\documentclass[11pt,a4paper]{article}
\usepackage{mathrsfs}
\usepackage{amsfonts}
\usepackage{epsf,epsfig,amsfonts,amsgen,amsmath,amstext,amsbsy,amsopn,amsthm}
\usepackage{cite}

\theoremstyle{definition}

\begin{document}
\bibliographystyle{plain}

\newcommand{\tabincell}[2]{\begin{array}{@{}#1{}}#2\end{array}}

\title{\bf {\ Optimal Ternary Linear Complementary Dual  Codes}}
%\date{2020.10.08 (Manuscript )}
\author{Liangdong Lu$^{a,\dag}$, Ruihu Li, Qiang Fu,  Chen Xuan, Wenping Ma$^{b}$\\
a. College of  Science,  Air Force Engineering University,  Xi'an,
 Shaanxi,\\ 710051, China, (email:
$^{\dag}$ kelinglv@163.com, )\\
  b.School of Telecommunications Engineering,
  Xidian University, Xi'an, \\Shaanxi 710071, China \\}
\maketitle

\begin{abstract}
Linear complementary dual (LCD) codes introduced by Massey are the
codes whose intersections with their dual codes are trivial. It can
help to improve the security of the information processed by
sensitive devices, especially against side-channel attacks (SCA) and
fault invasive attacks. In this paper, By construction of
puncturing, extending, shortening and combination codes, many good
ternary LCD codes are presented. We give a Table 1 with the values
of $d_{LCD}(n,k)$ for length $ n \leq 20$. In addition, Many of
these ternary LCD codes given in this paper are optimal which are
saturating the lower or upper bound of Grassl's codetable in
\cite{Grassl}  and some of them are nearly optimal.

\end{abstract}

\section{Introduction}
\noindent

Let $q$ be a  power of a prime $p$, $F_{q}$ be the finite field with
$q$ elements, and $F^{n}_{q}$ be the $n$-dimensional vector space
over $F_{q}$. A $q$-ary $[n,k,d]_{q}$ linear code over $F_{q}$ is a
$k$-dimensional subspace of $F^{n}_{q}$ with Hamming distance $d$.
For a given  $[n,k]_{q}$ linear code, the code
$\mathcal{C}$$^{\perp}=\{x\in F_{q}|x\cdot c=0, c\in \mathcal{C}\}$
is called the dual code of $\mathcal{C}$\cite{Macwilliams,Huffman}.
A matrix $G$ whose rows form a basis of $\mathcal{C}$ is called a
generator matrix of $\mathcal{C}$. The weight enumerator $W(z)$ of a
code $\mathcal{C}$ is given by $W(z)=\sum_{i=0}^{n}A_{i}z^{i}$,
where $A_{i}$ is the number of codewords of weight $i$ in
$\mathcal{C}$. If $\mathcal{C}=\mathcal{C}^{\bot}$, $\mathcal{C}$ is
called a self-dual code. A linear code $\mathcal{C}$ is formally
self-dual if $\mathcal{C}$ and $\mathcal{C}^{\bot}$ have the same
weight enumerator.

A $q$-ary linear code $\mathcal{C}$ is called a linear complementary
dual (LCD) code if it meets its dual trivially, that is
$\mathcal{C}\cap \mathcal{C}^{\perp} =\{\mathbf{0}\}$  and each
generator matrix $G$ of $\mathcal{C}$ must satisfy $k=$
rank($GG^{\dagger})$, which was introduced by Massey
\cite{Massey1964,Massey1992}.

Carlet and Guilley in \cite{Carlet2016} shown that  LCD codes play
an important role in armoring implementations against side-channel
attacks, and presented several constructions of LCD codes. In
addition, with their applications in data storage, communications
systems, and consumer electronics, LCD codes have been employed in
cryptography and quantum error correcting recently\cite{lai13}. To
determine the largest minimum weight among all $[n,k]_{q}$ LCD code
are fundamental problems.

Recently, Carlet, Mesnager, Tang, Qi, Pellikaan in \cite{Carlet2018}
shown that any  $[n, k, d]$-linear code over $F_{q^{2}}$ is
equivalent to  an $[n, k, d]$- linear Hermitian LCD code over $F
_{q^{2}}$ for $q > 2$. Araya, Harada and Saito in \cite{Araya1} give
some conditions on the nonexistence of quaternary Hermitian linear
complementary dual codes with large minimum weights. Inspired by
these works,   we study constructions of Ternary LCD code. Then some
families of Ternary LCD codes with good parameters are constructed
from the known optimal codes by puncturing, extending, shortening
and combination method. Compared with the tables of best known
linear codes (referred to as the {\it Database} later) maintained by
Markus Grassl at $http://$ www.codetables.de, some of our codes
presented in this paper are saturating the lower bound of Grassl's
codetable.

 If the minimum distance of $\mathcal{C}$ is $d$, then
$\mathcal{C}$ can be denoted as $\mathcal{C}$ $=[n,k,d]$. A code
 $\mathcal{C}$$=[n,k,d]$ is an {\it optimal} code if there is no
$[n,k,d+1]$ code. If  $d$ is the largest
 value present known that there  exists  an $[n,k,d]$, then
$\mathcal{C}$$=[n,k,d]$ is called a {\it best known }  code.
 Denote $d_{LCD}(n,k) = max \{d|$  an $[n,k,d]$ LCD code$\}$.
 If an $\mathcal{C}=[n, k, d_{LCD}(n,k)]$ LCD
code is saturating the lower or upper bound of Grassl's
codetable\cite{Grassl}, we call $\mathcal{C}$ an optimal LCD code
and denote $d_{LCD}(n,k)=d_{o}(n,k)$. If $d_{LCD}(n,k)=d_{o}(n,k)-1$
, we call $\mathcal{C}$ an nearly optimal LCD code.

All of the $[n,k,d]$ LCD codes for $k\leq 3$ constructed in this
paper are optimal. $[19,4,11]$, $[11,5,5]$, $[14,7,6]$, $[14,8,5]$
$[16,5,8]$, $[18,4,10]$, $[19,5,10]$, $[20,5,11]$ LCD codes are
optimal.

 This paper is organized
as follows. In Section 2 we provide some required basic knowledge on
Ternary  LCD codes and code constructions. We derive constructions
of Ternary LCD codes in Section 3. In Section 4, we discuss Ternary
LCD codes with good parameters.

\section{ Construction Methods}
\noindent In this section,  we introduce some  basic concepts on
ternary linear codes. Let $\mathbf{F}_{3}=\{0,1,2\}$ be the Galois
field with three elements. Denote the $n$-dimensional space over
$\mathbf{F}_{3}$ by $\mathbf{F}_{3}^{n}$, we call a $k$-dimensional
subspace $\mathcal{C}$ of $\mathbf{F}_{3}^{n}$ as an $k$-dimensional
linear code of length $n$ and denote it as $\mathcal{C}$
$=[n,k]_{3}$. In the following sections,
  an $[n,k,d]_{3}$ code is denoted as $[n,k,d]$
for short. Defining the  inner product of ${\bf u}$, ${\bf v}\in$
$\mathbf{F}_{3}^{n}$ as
  $({\bf u,v})={\bf uv}=u_{1}v_{1}+u_{2}v_{2}+\cdot\cdot\cdot+u_{n}v_{n}.$
 The dual code
of $\mathcal{C}$ $=[n,k]$ is $\mathcal{C}$$^{\perp h}$ $=\{x\mid
(x,y)_{h}=0,\forall y\in \mathcal{C}\}$, and $\mathcal{C}$$^{\perp
}$ $=[n,n-k]$.  A generator matrix $H=H_{(n-k)\times n}$ of
$\mathcal{C}$$^{\perp }$ is called a parity check matrix of
$\mathcal{C}$. If $\mathcal{C}$ $\subseteq$ $\mathcal{C}$$^{\perp
}$, $\mathcal{C}$ is called a {\it weakly self-orthogonal} code. If
$\mathcal{C}$ is a  self-orthogonal code then each generator matrix
$G$ of $\mathcal{C}$ must satisfy rank($GG^{\bot})=0$. If
$\mathcal{C}$ $\cap$ $\mathcal{C}$$^{\perp }=\{0\}$, then
$\mathcal{C}$ (or $\mathcal{C}$$^{\perp }$) is called  a {\it
 LCD} code,  see Refs.\cite{Massey1992}. If
$\mathcal{C}$ is a  self-orthogonal code then each generator matrix
$G$ of $\mathcal{C}$ must satisfy rank($GG^{\bot})=k$.

For LCD codes, some  characterization is as following
\cite{Massey1992}:

{\bf Theorem 2.1.} Let $G$ and $H$ be a generator matrix and a
parity-check matrix of a code $\mathcal{C}$ over $F_{q}$,
respectively. Then the following properties are equivalent:

(I)  $\mathcal{C}$ is a LCD code;

(II)  $\mathcal{C}^{\bot}$ is a LCD code;

(III)  $GG^{T}$ is nonsingular;

(IV)  $HH^{T}$ is nonsingular.

We give another important property in the following corollary:

{\bf Corollary 2.1.} The linear code $\mathcal{C}$ is LCD if and
only if $F_{q}^{n}=\mathcal{C}\oplus \mathcal{C}^{\bot}$.

In the following sections, by construction of puncturing, extending,
shortening and combination codes, we will discuss  ternary LCD code
$\mathcal{C}$ $=[n,k,d]$ with $d$ as large as possible for given
$n\leq 20$ or $k\leq 3$.
 Firstly, we
make some notations for later use.

Let $\bf{1_{n}}$=$(1,1,...,1)_{1\times n}$  and
$\bf{0_{n}}$=$(0,0,...,0)_{1\times n}$ denote the all-one vector
 and  the all-zero vector of length $n$, respectively.
Construct
$$S_{2}=\left(
\begin{array}{ccccc}
1&0&1&1\\
0&1&1&2\\
\end{array}
\right)=(\alpha_{1},...,\alpha_{4}),$$
 $$ S_{3}=
\left(
\begin{array}{cccccc}
S_{2}&\mathbf{0}_{2\times1}&S_{2}&S_{2}\\
\mathbf{0}_{4}&1&\mathbf{1}_{4}&2\cdot\mathbf{1}_{4}\\
\end{array}
\right)=(\beta_{1},\beta_{2},\cdots,\beta_{13}),$$
 $$S_{4}=
\left(
\begin{array}{cccccc}
S_{3}&\mathbf{0}_{3\times1}&S_{3}&S_{3}\\
\mathbf{0}_{13}&1&\mathbf{1}_{13}&2\cdot\mathbf{1}_{13}\\
\end{array}
\right)=(\gamma_{1},\gamma_{2},\cdots,\gamma_{40}).$$

$$\vdots$$

 $$S_{k}=
\left(
\begin{array}{cccccc}
S_{k-1}&\mathbf{0}_{k-1\times 1}&S_{k-1}&S_{k-1}\\
\mathbf{0}_{\frac{3^{(k-1)}-1}{2}}&1&\mathbf{1}_{\frac{3^{(k-1)}-1}{2}}&2\cdot\mathbf{1}_{\frac{3^{(k-1)}-1}{2}}\\
\end{array}
\right)$$

It is well known that the matrix $S_{2}$ generates the $[4,2,4]$
simplex code with weight polynomial $1+8y^{3}$,  $S_{3}$ generates
the $[13,3,9]$ simplex code with weight polynomial $1+26y^{9}$,
$S_{4}$ generates  the $[40,4,27]$ simplex code with weight
polynomial $1+80y^{27}$, $S_{5}$ generates  the $[121,5,81]$ simplex
code with weight polynomial $1+242y^{81}$, and
$S_{k}S_{k}^{\bot}=\mathbf{0}$ for $k=2,3,4,5,\cdots$, see
Ref.\cite{Macwilliams,Li2004}.

\section{ Bounds for $d_{LCD}(n,k)$ of Ternary LCD codes }
\noindent In this subsection, we construct optimal or near-optimal
 LCD codes over $F_{3}$. For $n\leq 10$, some optimal or
near-optimal ternary LCD codes are presented by Araya et al. in
Ref.\cite{Araya1}.

{\bf Case 1. Two  dimensional ternary LCD codes }

The parameters of two  dimensional optimal codes have been
determined as following Table 1 [16].
\begin{center}
Table 1. Parameters of optimal $[n,2]$  linear codes\\
\begin{tabular}{llllll}
  \hline
  % after \\: \hline or \cline{col1-col2} \cline{col3-col4} ...
  $n$ & 4s  & 4s+1 &4s+2&  4s+3  \\

  $d $& 3s & 3s &  3s+1 &  3s+2\\
  \hline
  \end{tabular}
\end{center}
The optimal $[4s,2,3s]$ and $[4s+3,2,3s+2]$ codes are not LCD codes
according to [16]. Let $G=G_{2,n}$ be a generate matrix of an
optimal
 $[n,2]$ code with $n=4s$ or $n=4s+3$, then rank($GG^{\top})\leq
 1$. Hence, the parameters of LCD $[n,2]$ codes for $n=4s$ and $n=4s+3$
maybe $[4s,2,3s-1]$ and $[4s+3,2,3s+1]$ respectively. Parameters of
optimal LCD $[n,2,d]$ codes are listed as following Table 2.

\begin{center}
Table 2. Parameters of  LCD $[n,2]$ codes\\
\begin{tabular}{llllll}
  \hline
  % after \\: \hline or \cline{col1-col2} \cline{col3-col4} ...
  $n$ & 4s  & 4s+1 &4s+2&4s+3\\

   $d $ & 3s-1 & 3s &  3s+1 &  3s+1\\
  \hline
  \end{tabular}
\end{center}

{\bf Theorem 3.1.}  Let $ n\geq 4$. Then

\quad\quad\quad\quad\quad\quad(i)
$d_{LCD}(n,2)=\lceil\frac{3n}{4}\rceil-1$ for $n\equiv 0,1 (mod 4)$;

\quad\quad\quad\quad\quad\quad(ii)
$d_{LCD}(n,2)=\lfloor\frac{2n}{3}\rfloor$ for $n\equiv 2,3 (mod 4)$.

{\bf Proof.}  For $n\geq 4s$ and $s\geq 1$,  construct
$G_{2,4s}=(\alpha_{1},2\alpha_{2},\alpha_{3}|(s-1)S_{2})$,
$G_{2,4s+1}=(2\alpha_{1},2\alpha_{2},\alpha_{3}|(s-1)S_{2})$,
$G_{2,4s+2}=(\alpha_{1},\alpha_{2}|(s-1)S_{2})$,
$G_{2,4s+3}=(2\alpha_{1},\alpha_{2}|(s-1)S_{2})$.

It is easy to check that: all of the codes with generator matrices
$G_{2,n}$ have the desired parameters as Table 2, and $2=$
rank($G_{2,n}G_{2,n}^{\dagger})$. Hence the lemma holds.

 {\bf Remark 3.1} According to Lemma 4 in \cite{Eupen}, a 3-ary code with parameters
$[4s-\epsilon, 2, 3s-\epsilon]$ is unique for $\epsilon=0,1$. Hence,
the  $[n,2]$ LCD code for $n=4s,4s-1$ in this case are optimal. The
$[n,2]$ LCD code for $n=4s+1,4s+2$  are optimal.

 {\bf Case 2. Three  dimensional ternary LCD codes}

In this case, we only discuss construction of three dimensional LCD
codes. We give parameters of optimal LCD $[n,3]$ codes listed in
Table 4 and Table 5. For $n=13s+t$, $s\geq 1$,  according to
classification of ternary codes in [chengang dalunwen], the optimal
$[13s,3,9s]$, $[13s+9,2,9s+6]$ and $[13s+12,2,9s+8]$ codes are not
LCD codes. Hence, the minimal distance of  $[n,3]$ LCD codes for
$n=13s$, $n=13s+9$ and $13s+12$ can not greater than $9s-1$, $9s+5$
and $9s+7$, respectively. The LCD $[n,3]$ code for $n=13s$ and
$s\geq 1$ has parameters $[13s,3, 9s-1]$. Parameters of good LCD
$[n,3,d]$ codes are listed as following Table 3 and Table 4.

\begin{center}
Table 3. Parameters of LCD $[n,3]$ codes for length $3\leq n\leq
13$\\[3mm]{

\begin{tabular}{llllllllllllllllllllll}
  \hline
  % after \\: \hline or \cline{col1-col2} \cline{col3-col4} ...
  $n$ & 3  & 4 & 5 & 6& 7 & 8 & 9 & 10&11&12& 13 \\

  $d$ & 1 & 2 &  2 & 3&  4& 4 & 5 & 6& 6 & 7& 8\\
  \hline
  \end{tabular}}
\end{center}
$\ $ \\

\begin{center}
 \noindent Table 4. Parameters of LCD $[n,3]$ codes for length $n\geq
14$\\ [3mm]{ $ \begin{tabular} {llllllllllllllllll}
\hline
$n$&$13s+1$&$13s+2$&$13s+3$&$13s+4$&$13s+5$&$13s+6$\\
 $d$&$9s-1$&$9s$ &$9s+1$ &$9s+1$
&$9s+2$&$9s+3$ \\
 \hline $n $&$13s+7$&$13s+8$&$13s+9$&$13s+10$&$13s+11$
&$13s+12$\\
 $d$&$9s+4$  &$9s+4$ &$9s+5$
 &$9s+6$&
$9s+6$&$9s+7$ \\
\hline
\end{tabular}$
}
\end{center}

 {\bf Theorem 3.2.} (1) If $3\leq n\leq 13$, then there exist $[n,3,d]$
 LCD codes as given in Table 3.\\
(2) If $ n=13s+t\geq 14$, then there exist $[n,3,d]$ LCD codes as
given in Table 4.

 {\bf Proof.} (1) The LCD codes are given in \cite{Araya1} for $4\leq n\leq
10$. Construct $G_{3,5}=(I_{3}|2\beta_{8})$,
$G_{3,6}=(I_{3}|\beta_{9},\beta_{11},\beta_{13})$,
$G_{3,7}=(I_{3}|\beta_{4},\beta_{7},\beta_{10},\beta_{13})$,
$G_{3,8}=(I_{3}|\beta_{7},\beta_{8},\beta_{10},\beta_{11},\beta_{12})$,
$G_{3,9}=(I_{3}|\beta_{3},\beta_{4},\beta_{6},\beta_{7},\beta_{11},\beta_{13})$,
$G_{3,10}=(G_{3,9}|\beta_{9})$,
 $G_{3,11}=(I_{3}|\beta_{2},\beta_{5},\beta_{6},\beta_{7},\beta_{8},\beta_{9},\beta_{10},\beta_{12})$,
 $G_{3,12}=(I_{3}|\beta_{3},2\beta_{8},2\beta_{9},2\beta_{10},\beta_{11},\beta_{12})$,
$G_{3,13}=(G_{3,12}|\beta_{13})$,
$G_{3,14}=(G_{3,12}|\beta_{2})$,\\
$G_{3,15}=(I_{3}|\beta_{1},\beta_{2},\beta_{3},\beta_{4},\beta_{6},\beta_{7},\beta_{8},2\beta_{10},2\alpha_{11},\alpha_{12})$,
$G_{3,16}=(G_{3,15}|\beta_{13})$, $G_{3,17}=(G_{3,16}|\beta_{9})$,

It is not difficult to check that rank($G_{3,n}G_{3,n}^{\dagger})=3$
for $4\leq n\leq 17$, and the codes $\mathcal{C}$$_{n}$ with
generator matrices
$G_{3,n}$ have weight polynomials  $W_{n}(z)$ as follows:\\
\begin{center}
$W_{3,5}(z)=$$1+6z^{2}+8z^{3}+6z^{4}+6z^{5}$,\\
 $W_{3,6}(z)=$$1+6z^{3}+12z^{4}+6z^{5}+2z^{6}$,\\
$W_{3,7}(z)=$$1+12z^{4}+6z^{5}+8z^{6}$,\\
$W_{3,8}(z)=$$1+2z^{4}+12z^{5}+8z^{6}+4z^{7}$,\\
 $W_{3,9}(z)=$$1+6z^{5}+8z^{6}+6z^{5}+12z^{7}$,\\
 $W_{3,10}(z)=$$1+8z^{6}+12z^{7}+6z^{8}$,\\
 $W_{3,11}(z)=$$1+6z^{6}+4z^{7}+12z^{8}+2z^{9}+2z^{10}$,\\
 $W_{3,12}(z)=$$1+10z^{7}+4z^{8}+8z^{9}+2z^{10}+2z^{11}$,\\
 $W_{3,13}(z)=$$1+12z^{8}+6z^{9}+6z^{10}+2z^{12}$,\\
$W_{3,14}(z)=$$1+4z^{8}+8z^{9}+10z^{10}+2z^{11}+2z^{13}$,\\
$W_{3,15}(z)=$$1+8z^{9}+4z^{10}+12z^{11}+2z^{13}$,\\
$W_{3,16}(z)=$$1+10z^{10}+6z^{11}+8z^{12}+2z^{13}$,\\
$W_{3,17}(z)=$$1+12z^{11}+8z^{12}+8z^{12}+6z^{13}$.
\end{center}

Summarizing the previous discussion, hence (1) holds.\\

(2) For $ n=13s+t\geq 14$. Construct
 $G_{3,13s}=(G_{3,13}\mid (s-1)
S_{3})$, $G_{3,13s+1}=(G_{3,14}\mid (s-1) S_{3})$,
 $G_{3,13s+2}=(G_{3,15}\mid (s-1) S_{3})$,
$G_{3,13s+3}=(G_{3,16}\mid (s-1) S_{3})$, $G_{3,21s+4}=(G_{3,17}\mid
(s-1) S_{3})$, $G_{3,21s+t}=(G_{3,t}\mid s S_{3})$ for $5\leq t\leq
12$.

 From  $ S_{3}S^{\dagger}_{3}=0$ and the
discussion of (1), one can deduce (2) holds.

 {\bf Remark 3.2} According to Lemma 6 in \cite{Eupen}, a 3-ary code with parameters
$[13-\epsilon, 3, 9-\epsilon]$ is unique for $0\leq\epsilon \leq 2$.
Hence, the  $[13s+\epsilon, 3, 9s-5+\epsilon]$ LCD code for
$11\leq\epsilon \leq 12$  in this case are optimal. And the
$[13s,3,9s-1]$ LCD code is optimal. Since $[13s+9,3,9s+6]$ code is
unique and self-orthogonal, $[13s+9,3,9s+5]$ LCD code code is
optimal.

{\bf Case 3. LCD codes of  length $11 \leq n\leq 20$ }

In this Case,  we only discuss construction of LCD codes of length
$11 \leq n\leq 20$. For $4\leq n\leq 10$, the LCD codes are given in
\cite{Araya1}.

 {\bf Lemma 3.3.} (1) $d_{LCD}(13,6)=6$, $d_{LCD}(14,7)=6$, $d_{LCD}(14,8)=5$, $d_{LCD}(15,6)=7$, $d_{LCD}(16,9)=5$,
   $d_{LCD}(19,12)=5$, $d_{LCD}(20,12)=6$, $d_{LCD}(20,13)=5$.\\
(2) $d_{LCD}(13,7)=5$, $d_{LCD}(14,6)=6$, $d_{LCD}(15,9)=4$,
$d_{LCD}(16,7)=6$, $d_{LCD}(19,7)=8$, $d_{LCD}(20,8)=8$,
$d_{LCD}(20,7)=8$.

 {\bf Proof.} (1) These LCD codes are listed in  Appendix.
The LCD codes in (2) are dual codes of ones in (1).

 {\bf Theorem 3.4.} (1) $d_{LCD}(14,5)=7$, $d_{LCD}(13,4)=7$,
 $d_{LCD}(13,5)=6$.

(2) $d_{LCD}(n,k)=6$ where  $17\leq n\leq 20$, $n-10\leq k\leq n-8$.

(3) $d_{LCD}(n,4)=\lceil \frac{11}{20}n\rceil$ where $19\leq n\leq
21$; $d_{LCD}(n,4)=\lceil \frac{9}{11}n\rceil$ where $15\leq n\leq
18$.

 {\bf Proof.}
Shortening  LCD code $[15, 6, 7]$ in Lemma 3.3 on coordinate sets
 $\{6 \}$ and $\{1,2 \}$, one can  obtain LCD codes $[14, 5, 7]$ and  $[13, 4, 7]$.
Puncturing  the $[14, 5, 7]$ on coordinate sets
 $\{1 \}$,  one can  obtain LCD code $[13, 5, 6]$.
\begin{center}
$W_{5,14}(z)=$$1+34z^{7}+56z^{8}+46z^{9}+36z^{10}+34z^{11}+34z^{12}+2z^{13}$.\\
$W_{4,13}(z)=$$1+16z^{7}+22z^{8}+20z^{9}+12z^{10}+8z^{11}+2z^{13}$.\\
$W_{5,13}(z)=$$1+18z^{6}+44z^{7}+56z^{8}+52z^{9}+28z^{10}+34z^{11}+10z^{12}$.\\
\end{center}

Construct

$A_{11,9}=\left(
\begin{array}{ccccc}
 1 1 0 0 1 1 1 1 0\\
 0 0 1 0 0 2 2 1 1\\
 0 2 1 1 1 0 0 2 0\\
 0 0 2 1 1 2 0 0 2\\
 0 1 2 2 0 2 2 0 1\\
 0 2 2 2 0 0 0 2 2\\
 0 1 1 2 1 0 1 0 0\\
 0 0 1 1 2 2 0 1 0\\
 0 0 0 1 1 1 1 0 1\\
 0 2 1 0 2 2 1 1 2\\
 0 1 1 1 2 1 2 1 2\\
\end{array}\right)$,
$A_{13,10}=\left(
\begin{array}{ccccc}
 0 0 0 0 1 0 1 1 1 1\\
 0 0 1 1 0 0 0 2 1 1\\
 0 0 1 0 2 1 0 1 2 1\\
 0 0 1 0 0 2 2 1 1 0\\
 0 0 0 2 0 0 1 2 1 2\\
 0 0 2 2 1 2 0 0 2 0\\
 0 0 0 1 1 2 1 0 0 1\\
 0 0 1 1 2 0 1 2 0 2 \\
 0 0 2 1 2 0 0 0 2 1\\
 0 0 1 2 2 2 0 1 0 0 \\
 0 0 0 2 1 1 1 0 1 0 \\
 0 0 0 0 1 2 2 1 0 2 \\
 0 0 2 2 0 1 1 1 1 1 \\
\end{array}\right)$,$A_{4,17}=\left(
\begin{array}{ccccc}
 1 2 2 1 0 0 2 1 2 0 1 0 2 2 2 0 0\\
 0 2 1 1 0 1 0 2 0 0 2 1 1 2 1 1 2\\
 1 2 1 2 1 0 0 1 1 0 1 2 0 0 1 1 1\\
 0 1 0 1 1 2 2 2 1 1 0 0 2 0 1 0 1\\
\end{array}\right)$,

Let $G_{11,20}=[I_{11}|A_{11,9}]$. Then there is a LCD code $[20,
11, 6]$  with generator matrix $G_{11,20}$.
$W_{11,20}(z)=$$1+314z^{6}+696z^{7}+1982z^{8}+4996z^{9}+10316z^{10}+17520z^{11}+25260z^{12}+30594z^{13}+
30804z^{14}+25354z^{15}+16968z^{16}+8422z^{17}+3124z^{18}+718z^{19}+78z^{20}$.

 Shortening $[20,11, 6]$ on coordinate sets
 $\{3\},\{2,11\} and \{1,2,4\}$, one can  obtain LCD code $[19, 10, 6]$,
 $[18, 9, 6]$ and $[17, 8, 6]$.

\begin{center}
$W_{10,19}(z)=$$1+204z^{6}+454z^{7}+1150z^{8}+2574z^{9}+4988z^{10}+7746z^{11}+9822z^{12}+
10734z^{13}+9462z^{14}+6588z^{15}+3548z^{16}+1406z^{17}+332z^{18}+40z^{19}$.\\

$W_{9,18}(z)=$$1+136z^{6}+264z^{7}+622z^{8}+1390z^{9}+2190z^{10}+3186z^{11}+3606z^{12}+
3414z^{13}+2670z^{14}+1406z^{15}+612z^{16}+164z^{17}+22z^{18}$.\\

$W_{8,17}(z)=$$1+128z^{6}+208z^{7}+502z^{8}+876z^{9}+1252z^{10}+1354z^{11}+1162z^{12}+
68z^{13}+304z^{14}+74z^{15}+12z^{16}$.\\
\end{center}

Let $G_{13,23}=[I_{13}|A_{13,10}]$. Then there is a $[23, 13, 6]$
code with generator matrix $G_{13,23}$. It is not a LCD code.
Shortening $[23, 13, 6]$ on coordinate sets
 $\{1,4,11\},\{ 3, 4, 9, 12 \}, \{ 1,8, 9, 12, 13\}$ and $\{ 4, 6, 8, 10, 11,
12\}$  ,  one can  obtain LCD code $[20, 10, 6]$,
 $[19, 9, 6]$, $[18, 8, 6]$ and $[17, 7, 6]$.

\begin{center}
$W_{10,20}(z)=$$1+324z^{6}+524z^{7}+1648z^{8}+3892z^{9}+6798z^{10}+
9906z^{11}+11610z^{12}+
10698z^{13}+7698z^{14}+3978z^{15}+1582z^{16}+350z^{17}+40z^{18}$.\\
$W_{9,19}(z)=$$1+212z^{6}+332z^{7}+930z^{8}+1886z^{9}+3012z^{10}+
3990z^{11}+3768z^{12}+
2970z^{13}+1704z^{14}+694z^{15}+166z^{16}+18z^{17}$.\\
$W_{8,18}(z)=$$1+132z^{6}+196z^{7}+514z^{8}+870z^{9}+1252z^{10}+
1366z^{11}+1144z^{12}+ 706z^{13}+280z^{14}+94z^{15}+ 6z^{16}$.\\
$W_{7,17}(z)=$$1+84z^{6}+110z^{7}+250z^{8}+380z^{9}+488z^{10}+
424z^{11}+258z^{12}+ 158z^{13}+28z^{14}+6z^{15}$.\\
\end{center}

Shortening $[20, 12, 6]$ in Lemma 3.3 on coordinate sets
 $\{3\},\{2,11\},\{8,9,10\}$, one can  obtain LCD code $[19, 11, 6]$,
 $[18, 10, 6]$, $[17, 9, 6]$.
\begin{center}
$W_{11,19}(z)=$$1+468z^{6}+840z^{7}+2882z^{8}+7284z^{9}+14408z^{10}+23646z^{11}+31296z^{12}+
34140z^{13}+28896z^{14}+19248z^{15}+9822z^{16}+3382z^{17}+752z^{18}+82z^{19}$.\\
$W_{10,18}(z)=$$1+316z^{6}+538z^{7}+1680z^{8}+3812z^{9}+6810z^{10}+9972z^{11}+11574z^{12}+
10740z^{13}+757z^{14}+4106z^{15}+1514z^{16}+372z^{17}+36z^{18}$.\\
$W_{9,17}(z)=$$1+212z^{6}+328z^{7}+928z^{8}+1910z^{9}+3012z^{10}+3948z^{11}+3774z^{12}+
2982z^{13}+1740z^{14}+664z^{15}+158z^{16}+26z^{17}$.
\end{center}
Shortening $[20, 8, 8]$ in Lemma 3.3 on coordinate sets
 $\{3,4\}, \{2,4,5\}$ and $\{1, 3, 7, 8 \}$, one can  obtain LCD code
 $[18, 6, 8]$,  $[17, 5, 8]$ and $[16, 4, 9]$.

\begin{center}
$W_{6,18}(z)=$$1+18z^{8}+48z^{9}+108z^{10}+150z^{11}+106z^{12}+
120z^{13}+84z^{14}+70z^{15}+24z^{16}$.\\
$W_{5,17}(z)=$$1+8z^{8}+28z^{9}+46z^{10}+58z^{11}+40z^{12}+
24z^{13}+24z^{14}+12z^{15}+2z^{16}$.\\
$W_{4,16}(z)=$$1+18z^{9}+18z^{10}+22z^{11}+12z^{12}+
6z^{13}+2z^{14}+2z^{15}$.
\end{center}

Let $G_{4,21}=[I_{4}|A_{4,17}]$. There is a $[21, 4, 12]$ LCD code
of generator matrix $G_{4,21}$ with weight enumerator
$1+12z^{12}+18z^{13}+20z^{14}+18z^{15}+4z^{16}+4z^{17}+2z^{18}+2z^{19}$.
Puncturing  the $[21, 4, 12]$ on coordinate sets
 $\{1 \}$, $\{7,16 \}$, $\{1, 2, 7 \}$,  $\{1, 2, 7,8 \}$ and $\{1, 2, 3,5,7,8 \}$   one can  obtain LCD code $[20, 4, 11]$, $[19, 4, 11]$,
 $[18, 4, 10]$,  $[17, 4, 9]$ and $[15, 4, 8]$.

\begin{center}
$W_{4,20}(z)=$$1+6z^{11}+18z^{12}+22z^{13}+14z^{14}+12z^{15}+2z^{16}+4z^{17}+2z^{18}$.\\
$W_{4,19}(z)=$$1+16z^{11}+24z^{12}+22z^{13}+6z^{14}+6z^{15}+2z^{16}+2z^{17}+2z^{18}$.\\
$W_{4,18}(z)=$$1+10z^{10}+18z^{11}+28z^{12}+12z^{13}+6z^{14}+2z^{15}+2z^{16}+2z^{18}$.\\
$W_{4,17}(z)=$$1+4z^{9}+20z^{10}+24z^{11}+14z^{12}+10z^{13}+4z^{14}+2z^{15}+2z^{17}$.\\
$W_{4,15}(z)=$$1+12z^{8}+20z^{9}+20z^{10}+12z^{11}+10z^{12}+4z^{13}+2z^{15}$.\\
\end{center}

 {\bf Theorem 3.5.} (1) $d_{LCD}(n,n-11)=8$ where $16\leq n\leq 17$; $d_{LCD}(n,n-10)=7$
where $15\leq n\leq 16$.

(2) $d_{LCD}(17,5)=9$, $d_{LCD}(18,5)=9$,
 $d_{LCD}(19,5)=10$.\\

 {\bf Proof.} Construct

$A_{6,11}=\left(
\begin{array}{ccccc}
 0 1 0 0 1 2 1 2 0 2 1\\
 1 1 1 2 1 1 2 0 2 0 2\\
 1 1 0 2 0 0 2 1 2 1 0\\
 1 2 2 2 0 1 1 2 1 1 1\\
 0 2 2 1 1 2 2 2 2 0 2\\
 1 1 2 0 2 0 1 0 0 2 2\\
\end{array}\right)$,$A_{5,12}=\left(
\begin{array}{ccccc}
 2 2 0 2 0 1 2 0 2 2 2 1\\
 2 1 2 1 1 1 1 2 0 0 0 2\\
 0 1 2 0 0 1 1 2 2 2 1 0\\
 2 1 2 0 2 2 0 1 2 1 2 2\\
 1 2 2 2 2 1 0 0 0 1 1 2\\
\end{array}\right)$,
 $G_{A,13}=\left(
\begin{array}{ccccc}
 2 2 2 2 2 2 2 2 0 0 0 0 0\\
 1 2 2 2 0 0 2 1 2 0 0 1 2\\
 2 2 1 0 0 2 1 0 2 1 0 2 0\\
 1 1 2 2 2 2 1 1 2 1 2 2 2\\
 2 0 0 1 2 0 0 2 0 1 1 1 2\\
\end{array}\right)$,
\begin{center}
$B=\left(
\begin{array}{ccccc}
1 0\\
2 1\\
2 0\\
2 0\\
0 0\\
\end{array}\right)$,$A_{5,15}=\left(
\begin{array}{ccccc}
 1 0 2 2 2 2 1 1 2 0 2 1 0 1 0\\
 0 1 0 2 2 0 2 1 2 2 0 2 2 0 1\\
 0 2 1 1 2 1 2 1 1 1 1 1 2 1 1\\
 1 2 1 1 0 0 1 2 2 0 2 0 0 2 2\\
 1 2 1 2 0 2 2 0 1 0 0 2 2 2 1\\
\end{array}\right)$,
$A_{9,11}=\left(
\begin{array}{ccccc}
 1 0 1 1 0 1 1 0 2 0 2\\
 2 1 0 2 1 1 1 0 2 0 1\\
 1 2 1 2 2 0 1 2 1 1 2\\
 2 1 2 2 2 0 0 0 1 2 2\\
 2 2 1 0 2 0 0 2 2 2 0\\
 0 2 2 1 0 2 0 0 2 2 2\\
 2 0 2 0 1 1 2 2 2 0 0\\
 0 2 0 2 0 1 1 2 2 2 0\\
 0 0 2 0 2 0 1 1 2 2 2\\
\end{array}\right)$,
\end{center}

Let $G_{6,17}=[I_{6}|A_{6,11}]$.There is a $[17, 6, 8]$ LCD code of
generator matrix $G_{6,17}$ with weight enumerator
$1+52z^{8}+82z^{9}+124z^{10}+136z^{11}+110z^{12}+124z^{13}+64z^{14}+32z^{15}+4z^{16}$.
Puncturing  the $[17, 6, 8]$ on coordinate sets
 $\{1 \}$,    one can  obtain LCD code $[16, 6, 7]$.
Shortening $[17, 6, 8]$ on coordinate sets
 $\{2 \}$,  one can  obtain LCD code $[16, 5, 8]$.
 Puncturing  the $[16, 5, 8]$ on coordinate sets
 $\{1 \}$,    one can  obtain LCD code $[15, 5, 7]$.
 Let $G_{5,17}=[I_{5}|A_{5,12}]$, $G_{5,18}=[I_{5}|A_{5,13}]$,
$G_{5,19}=[G_{5,17}|B]$and $G_{5,20}=[I_{5}|A_{5,15}]$. There are
LCD codes $[17, 5, 9]$, $[18, 5, 9]$, $[19, 5, 10]$ and $[20, 5,
11]$ of generator matrix $G_{5,17}$, $G_{5,18}$, $G_{5,19}$ and
$G_{5,20}$, respectively.

\begin{center}
$W_{6,16}(z)=$$1+24z^{7}+76z^{8}+102z^{9}+140z^{10}+136z^{11}+118z^{12}+86z^{13}+40z^{14}+4z^{15}+2z^{16}$.\\
$W_{5,16}(z)=$$1+20z^{8}+44z^{9}+64z^{10}+42z^{11}+28z^{12}+26z^{13}+10z^{14}+8z^{15}$.\\
$W_{5,15}(z)=$$1+6z^{7}+46z^{8}+52z^{9}+50z^{10}+36z^{11}+28z^{12}+16z^{13}+8z^{14}$.\\
\end{center}

\begin{center}
$W_{5,17}(z)=$$1+34z^{9}+60z^{10}+48z^{11}+36z^{12}+28z^{13}+22z^{14}+10z^{15}+2z^{16}+2z^{17}$.\\
$W_{5,18}(z)=$$1+20z^{9}+36z^{10}+48z^{11}+36z^{12}+28z^{13}+22z^{14}+10z^{15}+2z^{16}+2z^{17}$.\\
$W_{5,19}(z)=$$1+34z^{10}+44z^{11}+46z^{12}+40z^{13}+24z^{14}+32z^{15}+14z^{16}+4z^{17}+2z^{18}+2z^{19}$.\\
$W_{5,20}(z)=$$1+48z^{11}+44z^{12}+50z^{13}+32z^{14}+28z^{15}+22z^{16}+10z^{17}+8z^{18}$.\\
\end{center}

Let $G_{9,20}=[I_{9}|A_{9,11}]$.There is a $[20, 9, 8]$ LCD code of
generator matrix $G_{9,20}$ with weight enumerator
$1+390z^{8}+520z^{9}+3840z^{11}+2880z^{12}+64z^{14}+32z^{15}+4z^{16}$.

 {\bf Theorem 3.6.} (1) $d_{LCD}(n,n-4)=3$ where $11\leq n\leq 20$;
 $d_{LCD}(n,n-5)=4$ where $11\leq n\leq 16$; $d_{LCD}(n,n-5)=3$ where $17\leq n\leq
 20$;
 $d_{LCD}(n,n-7)=5$
where $15\leq n\leq 20$.

(2) $d_{LCD}(17,5)=9$, $d_{LCD}(18,5)=9$,
 $d_{LCD}(19,5)=10$.\\

 {\bf Proof.}
The dual code of $[20, 5, 11]$ is $[20, 15, 3]$ LCD code. Shortening
the $[20, 15, 3]$ on coordinate sets
 $\{10 \}$, $\{9,13 \}$ and $\{7,9,12 \}$,     one can  obtain LCD code $[19, 14, 3]$,$[18, 13, 3]$,$[17, 12, 3]$.

{\bf Theorem 3.7.}  Let $ n\geq 5$. Then

(i) $d_{LCD}(n,1)=n$ for $n\not\equiv 0($mod   $3)$; and
$d_{LCD}(n,n-1)=2$ for $n\not\equiv 0($mod  $3)$.

(ii) $d_{LCD}(n,1)=n-1$ for $n\equiv 0($mod  $3)$; and
$d_{LCD}(n,n-1)=1$ for $n\equiv 0($mod  $3)$.

(iii) $d_{LCD}(n,n-2)=2$.

 {\bf Proof.} Let $G_{1\times n}=\left(
\begin{array}{cccccccc}
a_{1}, a_{2}, \cdots, a_{i}, \cdots, a_{n}\\
\end{array}\right)$ be a generater matrix of code,  where $a_{i}\in F_{3}$.
According to the finite field theory, if $d_{LCD}(n,1)=n$, then
$a_{i}\neq 0$. For $n\not\equiv 0($mod   $3)$, $G_{1\times
n}G_{1\times n}^{\top}=\sum a_{i}^{2}=1$; For $n\equiv 0($mod $3)$,
$G_{1\times n}G_{1\times n}^{\top}=\sum a_{i}^{2}=0$.

Let
 $$H_{n-1\times n}=\left(
\begin{array}{cccccccc}
1& 0& \cdots  & 0 & b_{1}\\
0 &1&  \cdots & 0 & b_{2} \\
 & &  \ddots &    & \\
0 &0  & \cdots &1  & b_{n-1}\\
\end{array}\right)$$
 be a generater matrix of code,  where $b_{i}\in
F_{3}$. According to the finite field theory, if $d_{LCD}(n,n-1)=2$,
then $b_{i}\neq 0$. For $n\not\equiv 0($mod   $3)$,
$Rank(H_{n-1\times n}H_{n-1\times n}^{\top})=n-1$; For $n\equiv
0($mod $3)$, $Rank(H_{n-1\times n}H_{n-1\times n}^{\top})=n-2$;

Hence, (i) and (ii) hold.

\section{Conclusion}

In this paper, we  study  constructions of  ternary LCD codes. For
each $k\leq 3$ or $n\leq 20$,  we try all possible coordinates and
chose results in ternary LCD $[n,k]$ code with great minimum
distance as output. According to weight enumerators for
classification codes in \cite{Eupen} and codetable in \cite{Grassl},
all of the $[n,2,d]$ LCD codes constructed in this paper are
optimal. The  $[13s+\epsilon, 3, 9s-5+\epsilon]$ LCD code for
$11\leq\epsilon \leq 12$  in this case are optimal. And the
$[13s,3,9s-1]$ and  $[13s+9,3,9s+5]$ LCD codes are optimal.
According to weight enumerators for classification codes in
\cite{Eupen}, the number of classes of $[19,4,12]$, $[11,5,6]$,
$[11,6,5]$, $[12,6,6]$, $[16,5,9]$  is one and  not LCD, hence
$[19,4,11]$, $[11,5,5]$, $[16,5,8]$  LCD codes are optimal. And the
two number of classes of optimal $[18,4,11]$, $[19,5,11]$,
$[20,5,12]$  codes are all not LCD codes, hence, $[18,4,10]$,
$[19,5,10]$ $[20,5,11]$ LCD code are optimal. According to the
codetable in \cite{Grassl}, $[14,7,6]$, $[14,8,5]$ LCD codes  are
optimal.

Except these mentioned above codes, some of these ternary LCD code
constructed in this paper are optimal codes and saturate the bound
on  the minimum distance of codetable in \cite{Grassl}, and some of
them constructed in this paper do not attain known upper or lower
bounds on the minimum distance of a linear code. Nonetheless, the
minimum distances of those codes still seems to be very good in
general. In other words, these codes are the best possible among
those obtainable by our approach.

In Table 1, from puncturing, extending, shortening and combination
codes construction, many lower and upper bounds on minimal distance
of ternary LCD codes with length $n\geq 20$ are listed . In order to
make the bounds in Table 1 tighter, we need to choose other ternary
LCD codes better than that given in this paper and investigate other
code constructions to raise the lower bounds. We also plan to
explore the construction of ternary LCD codes from geometric view to
decrease the upper bound.

\begin{center}
 \noindent Table 1. Lower and upper bounds on the minimum distance of
    Ternary LCD codes.  The bold face entries represent improvements over the prior works.\\
     [3mm]{ $ \begin{tabular}
%{l|l|l|l|l|l|l|l|l|l|l|l|l|l|l|l|l|l|l|l|l|l|l|l|}
{llllllllllllllllllllllll} \hline $\scriptstyle{n\setminus k}$

    &1 &2  &3     &4        &5        &6        &7         &8      &9      &10    &11     &12        \\
%\hline
3  &2  &  &       &         &         &          &        &        &       &      &       &      \\
4  &4  &2  &      &         &         &          &        &        &       &      &       &     \\
5  &5  &3  &2     &2        &         &          &        &        &       &      &       &     \\
6  &5  &4  &3     &2        &1        &          &        &        &       &      &       &     \\
7  &7  &4  &4     &3        &2        &2         &        &        &       &      &       &     \\
8  &8  &5  &4     &4        &3        &2         &2       &        &       &      &       &     \\
9  &8  &6  &5     &4        &3        &3         &2       &1       &       &      &       &     \\
10 &10 &7  &6     &5         &4        &3        &3       &2       &2      &     &       &    \\
11 &11 &7  &6     &6        &5        &4         &3       &2        &2      &2     &      &     \\
12 &11 &8   &7    &6        &6         &5       &4       &3       & 2      &2     &1      &    \\

13 &13 &9 &8     &7       &6         &6         &5       &4       &3      &2      &2      &2     \\
14 &13 &10 &8     &8      &7         &6         &6       &5       &4      &3     &2       &2     \\
15 &15 &10 &9    &8       &7         &7         &6       &5       &4      &4     &3      &2     \\
16 &15 &11 &10    &9      &8         &7         &6       &6       &5      &4     &4      &3      \\
17 &17 &12 &11    &9      &9         &8         &6-7     &6-7     &6-7    &5     &4    &3            \\
18 &17 &13 &11    &10     &9         &8-9       &7-8        &6-7     &6-7    &6    &5    &4        \\
19 &19 &13 &12    &11     &10        &8-9        &8-9     &7-8        &6-7    &6     &6    &5      \\
20 &19 &14 &13    &11     &11        &9-10        &8-9     &8-9     &7-8       &6     &6    &6         \\

\hline
\end{tabular}$
}
\end{center}

\begin{center}
  { $ \begin{tabular}
{lllllllllllllllllllllllll} \hline $\scriptstyle{n\setminus k}$

             &13    &14       &15        &16       &17       &18     &19          \\
\hline
14           &2     &         &          &         &         &       &         \\
15           &2     &1        &          &         &         &       &         \\
16           &2     &2        &2         &         &         &       &         \\
17           &3    &2        &2         &2        &         &       &         \\
18           &3     &3        &2         &2        &1        &       &          \\
19           &4     &3        &3         &2        &2        &2      &          \\
20           &5    &4        &3         &3        &2        &2      &2           \\

\hline
\end{tabular}$
}
\end{center}
$\\$

\section*{Acknowledgements}
This work is supported by the National Natural Science Foundation of
China under Grant Nos.11801564 and 11901579.

\bibliography{wenxian}

\end{document}